\documentstyle[psfig,prb,aps]{revtex}
\begin{document}
\draft

\twocolumn[\hsize\textwidth\columnwidth\hsize\csname
@twocolumnfalse\endcsname

\title{
Ab initio calculations of structures and stabilities of
(NaI)$_n$Na$^+$ and (CsI)$_n$Cs$^+$ cluster ions.
}
\author{Andr\'es Aguado}
\address{Departamento de F\'\i sica Te\'orica, Facultad de Ciencias,
Universidad de Valladolid, 47011 Valladolid, Spain}
\author{Andr\'es Ayuela}
\address{Laboratory of Physics, Helsinki University of Technology,
02015 Espoo, Finland}
\author{Jos\'e M. L\'opez and Julio A. Alonso}
\address{Departamento de F\'\i sica Te\'orica, Facultad de Ciencias,
Universidad de Valladolid, 47011 Valladolid, Spain}
%\date{submitted to Phys.~Rev.~B, July, 11, 1997}
\maketitle
\begin{abstract}
{\em Ab initio} calculations using the Perturbed Ion model, 
with correlation contributions
included, are presented for 
nonstoichiometric (NaI)$_n$Na$^+$ and (CsI)$_n$Cs$^+$
(n$\le$14) cluster ions. The ground state and several low-lying isomers are
identified and described.
Rocksalt ground states are common and appear at cluster sizes
lower than in the corresponding neutral systems. 
The most salient features of the measured mobilities seem to be explained by
arguments related to the changes of the compactness of the clusters as a
function of size.
The stability of the
cluster ions against evaporation of a single alkali halide molecule
shows variations that explain the enhanced stabilities found experimentally
for cluster sizes n=4, 6, 9, and 13.
Finally, the ionization energies and the orbital eigenvalue spectrum of two
(NaI)$_{13}$Na$^+$ isomers are calculated 
and shown to be a fingerprint of the structure.
\end{abstract}

\vskip2pc]

\section{Introduction}

Alkali halide clusters have received substantial experimental and theoretical
attention. From the experimental side, they are relatively easy to form, and
their ionic bonding character has allowed the development of
simple models for the interaction between the
ions that form the cluster. They provide a nice opportunity to study the
emergence of condensed matter properties. An
important point in this respect is the determination of the most stable
isomers for each cluster size.

In this paper, we present the results of {\it ab initio} calculations of the
structures and stabilities of (NaI)$_n$Na$^+$ and (CsI)$_n$Cs$^+$ clusters
with n=1-14.
Large clusters of these materials were first produced by
bombardment of crystalline targets with Xe$^+$ ions and detected 
by means of secondary-ion
mass spectrometry.~\cite{Bar81,Cam81,Bar82} The mass spectra showed anomalies
in the cluster intensities for certain values of n, which were tentatively
interpreted as revealing the enhanced stability of ``cuboid-like'' structures.
The results were explained in terms of a direct emission model, which
assumes that ``cuboid-like'' cluster ions are directly sputtered from the
crystal. Ens {\em et al.}~\cite{Ens83} performed time-of-flight measurements
for (CsI)$_n$Cs$^+$ clusters, again produced by bombardment of CsI crystals, and
considered different observation times after cluster emission. In that way,
they found that the anomalies in the mass spectra were a consequence of
evaporative decay {\it after} production of the cluster ions, and that the
preferred decay channel was the ejection of a CsI molecule. Anomalies in the
mass spectra of clusters formed by the inert-gas condensation technique
were observed by Pflaum {\em et al.}~\cite{Pfl86}.
Twu {\em et al.}
published mass spectra of sodium chloride, sodium iodide, cesium
chloride, and cesium iodide cluster ions~\cite{Twu90} produced by
laser vaporization.
They observed the same magic numbers
for all those materials, namely n=4, 6, 9, and  13 in the small
size range. The differences in the detailed structure of the mass spectra
were attributed to the relative sizes of the ions making up the clusters. All
those techniques, however, give no direct information of the cluster
shapes. Drift cell experiments, which measure the mobility of the cluster ions
and stand as a promising technique for structural analysis of clusters,
have been performed to study the structures of small covalent and metallic
clusters.~\cite{Hel91,Cle93,Hun94,Jar95} More recently, ion mobility
experiments have been performed for alkali halide 
clusters~\cite{Mai97} and have revealed isomerization 
transitions~\cite{Hud97,Doy98} in (NaCl)$_n$Cl$^-$.

It was pointed out by Martin~\cite{Mar83} that precise measurements of 
the photoionization spectrum should also help
in determining the structure of clusters of ionic materials. The reason is that,
due to the strong localization of the electrons in closed-shell
alkali halide clusters, the structure of the photoionization spectrum should
be a fingerprint of the structure of the cluster, giving information on the
set of inequivalent anions.
Li {\em et al.} used this idea
to correlate the optical absorption spectra of (CsI)$_n$Cs$^+$
with the structures obtained by using a pair potential
model.~\cite{Li92,Li93a,Li93b} Photoelectron and photoionization
spectroscopy as well as theoretical studies have also been
performed to study the relation between the cluster structure and the localization
mode for small alkali halide clusters with 
excess electrons,~\cite{Yan92,Dur97,Fat96a,Fat96b}
and to study the structure and
emergence of metallic properties in alkali-rich alkali halide and
alkali hydride clusters.~\cite{Yu95,Bon96,Ant97,Fat97,Fra97}

Some theoretical calculations for alkali halide clusters have been based on
phenomenological pair potential models.~\cite{Mar83,Li93b,Wel76,Die85,Phi91}
Such simplified models have been successful in explaining their main 
characteristics, and furthermore, are very useful to find
the different local minima in the potential energy surface.
{\it Ab initio} calculations performed by
Ochsenfeld {\em et al.}~\cite{Och94,Och95a,Och95b} used molecular orbital
methods including correlation at the MP2 level.
We have used the
{\em ab initio} perturbed ion (PI) model,~\cite{Lua90a} which is formulated
within the restricted Hartree-Fock (RHF) 
approximation, in
studies of neutral stoichiometric alkali halide~\cite{Ayuel,Agu97a,Agu97c}
and (MgO)$_n$ clusters,~\cite{Agu97b} 
and in a preliminary study of nonstoichiometric (NaCl)$_n$Na$^+$
clusters.~\cite{Ayu97} In some of those studies correlation contributions were
included using an unrelaxed Coulomb-Hartree-Fock (uCHF) model proposed by
Clementi.~\cite{Cle65,Cha89}
The PI model represents a
major advance with respect to pair potential methods
and provides an alternative description to the molecular orbital
methods. In this paper we present an extensive study of charged (NaI)$_n$Na$^+$
and (CsI)$_n$Cs$^+$ clusters with n up to 14.
The results are aimed to assist in the
interpretation of the experimental investigations of the structure 
of alkali halide clusters, as provided by the ion mobility studies, or of
the relation between structure and photoionization spectrum.

The remainder of this paper is organized as follows:
In Sec.~\ref{sec:theory} we briefly review the computational method used in this
study. 
In Sec.~\ref{sec:results} 
we report the results for the isomer structures, relative stabilities, and
ionization potentials. 
Section~\ref{sec:summary} summarizes our
conclusions.

\section{Computational Method}
\label{sec:theory}

The {\em ab initio} perturbed ion model,\cite{Lua90a} as adapted to the study
of ionic clusters, has been described at length in our previous 
works.~\cite{Ayuel,Agu97a} In brief, the HF equations of the
cluster are selfconsistently solved in localized Fock spaces, 
by breaking the cluster wave
function into local nearly orthogonal ionic wave functions
and accounting for ion-cluster consistency.
The average binding energy per ion of the
(AX)$_n$A$^+$ cluster is given by
\begin{equation}
\label{binding:energy}
E_{\rm bind}=\frac{1}{2n+1}[nE_0(X^{-}) + (n+1)E_0(A^{+}) - E_{\rm cluster} ],
\end{equation}
where E$_0$(A$^{+}$) and E$_0$(X$^{-}$) are the energies of the free cation A$^{+}$
and anion X$^{-}$, respectively.
The localized nature of the PI model wave functions has some advantages.
In weakly overlapping systems, the correlation 
energy is almost intra-atomic.
In this 
paper, the correlation energy correction is obtained through Clementi's
Coulomb-Hartree-Fock method.~\cite{Cle65,Cha89}
The PI model also allows for the development of efficient computational
codes~\cite{Lua93} which make use of large multi-zeta Slater-type basis
sets~\cite{Cle74,Mcl81} for the description of the ions. Our
calculations have been performed using the following basis sets:
(11s9p5d) for Cs$^{+}$, taken from Ref.~\onlinecite{Mcl81};
(5s4p) for Na$^+$, and (11s9p5d) for I$^{-}$, both taken from 
Ref.~\onlinecite{Cle74}.
As input geometries for the optimization of the atomic structure
we have considered the structures obtained from pair
potential calculations.~\cite{Mar83,Li93b,Die85,Phi91}
Those input geometries have been
fully relaxed 
for (NaI)$_n$Na$^+$ (n $\le$ 6) and (CsI)$_n$Cs$^+$
(n $\le$ 4) clusters,
that is, the total cluster energy has been minimized with 
respect to variations in all the (3N-6) independent coordinates, where N
is the number of ions.
For larger clusters, full relaxations are 
expensive and we have relaxed these structures 
with respect to a limited number of relevant parameters, which depend on the
spatial symmetry of each isomer.
A downhill simplex algorithm has been used in all the 
optimizations.~\cite{Wil91}
All the cluster energies are converged within a precision of 1 meV.

\section{Results}
\label{sec:results}

\subsection{Structures of Isomers}

The calculated structures of small (NaI)$_n$Na$^+$ and (CsI)$_n$Cs$^+$
clusters
are shown in Figs. 1 and 2,
respectively. Small spheres represent cations and large spheres 
represent
anions.
The most stable structure, or ground state (GS)
is shown on the left side for each cluster size. The other structures
represent low-lying isomers.
Below each isomer, the energy difference with respect to the
ground state is given.
The structures
exhibit several distinct motives which can be roughly classified in
chains, rocksalt pieces, and rings (mainly hexagonal). 
It is possible that other isomers could exist between the ground state and
the low-lying isomers plotted in the figure, since our search has been
limited mainly to the structures provided by pair potential 
calculations,~\cite{Mar83,Li93b,Die85,Phi91} and the 
possibility of overlooking some isomers can not be
excluded. For n $\ge$ 10, the
rocksalt motives consistently dominate the ground state 
and the crystalline structure emerges, although signs of the appearance of the
rock-salt structure are also found for some of the clusters with $n < 10$. Let
us describe first the structures obtained for the smaller cluster sizes.
A linear chain is obtained as the ground state of (NaI)$_2$Na$^+$.
Bent chains, a rhombus with a cation attached, and a three-dimensional isomer
are found less stable. All those structures are closer
in energy for (CsI)$_2$Cs$^+$ and its three-dimensional GS is degenerate with
the linear chain.
The same
GS structure is found for both materials at n=3, namely a
cube with an anion removed. The linear chain is still the second isomer
for NaI, while it is found at a higher energy in CsI. A planar
rocksalt piece plus a cation appears as a high lying isomer. 
The rest of the isomers are obtained by
attaching a molecule in several ways to some of the n=2 structures. The n=4 GS
can be described as a quasi-two-dimensional sheet, which is
quite curved in NaI but almost planar for CsI. The chain isomers are not
competitive anymore.
A cube with a cation attached is
obtained as the second isomer. Again an essentially two-dimensional sheet, 
derived from the GS structure of (NaI)$_4$Na$^+$ is the GS of
(NaI)$_5$Na$^+$. This isomer is still more stable than the 
three-dimensional (3D) structures,
although the $2\times 2\times 3$
rocksalt piece with a corner anion removed is energetically very close. 
This notation indicates the number of ions along the x, y and z directions,
respectively.
For CsI,
a cube with a linear chain on-top is obtained as the GS.
The tendency for three-dimensional structures is stronger for the Cs-clusters.
Rounded cages with quadrangular and hexagonal faces form the GS for n=6 and n=7 in Fig. 1.
Rocksalt isomers are still high in energy, although an hexagonal isomer is 
only 0.02 eV above the GS in (NaI)$_6$Na$^+$.
GS structures with higher coordination are
obtained for CsI at these cluster sizes: a centered hexagonal prism
for n=6, and a complex structure containing a cube
for n=7. 
We can notice that the centered hexagonal prysm is also related to the
rock-salt structure since the central cation is coordinated to six ions.
Both in the ground state and in the low-lying isomers discussed until
now (n $\le$ 7) we appreciate a tendency to distorted structures that, in the
more extreme cases leads to elongated clusters. This is driven by the excess
positive charge. The distortion (elongation) lowers the repulsion between
cations.

A $3\times 3\times 2$-like structure with an anion missing from the
center of a basal plane is obtained as
the GS for n=8. The absence of this cation induces a distortion of that basis
to an octogonal ring.
A structure with an eight-coordinated cation is
close to the GS for CsI. Eight is the coordination number of
solid CsI. Structures obtained by adding a cation in several ways to the
hexagonal prismatic form of (NaI)$_9$ (which is the GS of the neutral 
cluster \cite{Agu97a}) are the most stable (NaI)$_9$Na$^+$ 
isomers. In the GS, this cation induces a strong cluster deformation.
The (CsI)$_9$Cs$^+$ GS is also related to an hexagonal prysm.
Cuboid-like rocksalt structures are 
less stable in both systems, but we can again notice that the upper half of
the hexagonal prysm is distorted in (CsI)$_9$Cs$^+$, and contains an inner
cation with coordination six (the rock-salt coordination). 
Starting with n=10 the emergence of cuboid-like rocksalt
features becomes apparent. From n=10 to n=14, all the NaI GS clusters
have the rocksalt structure. Notice in particular the high stability for n=13.
More open structures with lower
average coordination, or based on a stacking of rings are less stable.
The same can be said of CsI, with due exception of n=12,
where a hexagonal prismatic structure (although with a central
six-coordinated cation) is
more stable.

The structural trends of neutral alkali halide clusters have been studied
in Ref. \onlinecite{Agu97c}. Those trends
were rationalized in
terms of the relative ion sizes. 
As the ratio r$_C$/r$_A$ between the cation and anion radii increases, the
tendency to form rocksalt fragments becomes enhanced. NaI has a small value of
this ratio (0.44) while CsI has a large ratio (0.76). In spite of this
difference, the charged nonstoichiometric clusters behave in a similar way.
Rocksalt pieces appear early; about three quarters of the clusters between n=3
and n=14 have a rocksalt fragment, or a closely related structure, as the GS.
The exceptions occur for n=6, 7, 9; in these cases 
the rocksalt isomers have one or two 
low-coordinated cations. One can notice the influence of nonstoichiometry
and net charge: the percent of GS rocksalt structures in neutral (NaI)$_n$
is only about one half.~\cite{Agu97a} The reason seems to be that hexagonal
prysmatic structures are less competitive for the charged
nonstoichiometric clusters. A perfect prysm
formed by hexagonal rings has equal number of cations and anions, so only
defect-structures, obtained by removing an anion or by adding a cation, can be
built for nonstoichiometric clusters. 
We find one example of the first type in one of the isomers of
(NaI)$_5$Na$^+$. A cation can be added on top of a terminal ring or in the
interstitial hole between two hexagonal rings: isomers of both types exist for
(NaI)$_6$Na$^+$. These only become competitive when the rocksalt isomers are
very unstable, like for n=6, 9.

Pair potential calculations have been performed by
Diefenbach and Martin \cite{Die85} and Li {\em et al.} \cite{Li93b} 
for these two systems.
A comparison with the PI results
shows some discrepancies. The pair potential calculations for (NaI)$_n$Na$^+$
with n=8, 10, 11, 12, and 14 predict rather complex GS structures
(independently of the use or not of polarization terms) which correspond to
some of the isomers in fig. 1, whereas the PI calculations lead to
rock-salt-type structures.
For (CsI)$_n$Cs$^+$ the discrepancies are rather insignificant.
In the PI model the binding energy of the cluster can be written \cite{Agu97a}
as a sum of classical and quantum interaction energies between the ions plus a
term that accounts for the radial deformation of the electronic cloud of the
free ions (in practice the anions) in response to the environment. These
energy contributions contain quantum many electron terms that in principle
describe the interatomic interactions better than the empirical potentials.
There is also a second type of discrepancies between the PI and pair
potential calculations, although much less significant. These occur when the
rigid-ion and polarizable-ion model potentials disagree with each other and the
PI model agrees with one of them. These cases are found for (NaI)$_n$Na$^+$
with n=2, 5, and for (CsI)$_n$Cs$^+$ with n=2, 7, 8, 11, 12. In those cases
the PI model agrees sometimes with the rigid ion model and sometimes with the
polarizable ion model, but the two isomers involved are generally close in
energy, so the nonuniformity of the agreement can be ascribed to the small
energy differences involved. 
In addition one has to bear in mind that, due to
one basic assumption of the PI model (spherically symmetric electron density
clouds, centered on the nuclei), the polarization contribution coming from
dipolar terms is absent in this model.

Kreisle and coworkers \cite{Mai97} have studied the mobility of (NaI)$_n$Na$^+$
and (CsI)$_n$Cs$^+$ clusters under the influence of an electric field in a
chamber filled with helium gas. In these experiments the mobility is larger
the lower the He scattering cross section by the cluster, and this cross
section is inversely related to the compactness of the cluster. Kreisle and
coworkers have plotted the inverse mobilities as a function of cluster
size. Some salient features are
common to the two curves and, in our opinion, can be related to the structural
features found in Figs. 1 and 2. The main feature is a clear drop in the 
inverse mobility between n=12 and n=13. In fact, the inverse mobility becomes
a local minimum for (NaI)$_{13}$Na$^+$. It is suggestive to associate the high
mobility of n=13 to its compact ``perfect cube'' form. Other feature is a
visible change (decrease) of the slope of the inverse mobility curve at n=4.
It is tentative to associate this to the change from the two-dimensional to
more compact three-dimansional character of the ground state of (CsI)$_n$Cs$^+$
between n=4 and n=5. Although the calculated GS of (NaI)$_5$Na$^+$ is planar, there
is a low-lying isomer, only 0.02 eV higher in energy, that could easily be present in
the beam and contribute to increase
the mobility.
 
\subsection{Relative Stabilities as a Function of Cluster Size
and Comparison with Experiment}

The experimental mass spectra of alkali halide cluster ions
\cite{Bar81,Cam81,Bar82,Ens83,Pfl86,Twu90} show intensity
anomalies which reflect the special stability of some cluster sizes.
In order to study the relative stability of (NaI)$_n$Na$^+$ and (CsI)$_n$Cs$^+$
cluster ions, we plot in Fig. 3 the average binding energies per ion
(eq. 1) as a function of n. Maxima or pronounced changes of slope
in these curves are considered as indicators of enhanced stability.
Clear maxima at n=4 and n=13, and main
slope changes at n=6 and n=10, are obtained for (NaI)$_n$Na$^+$.
For (CsI)$_n$Cs$^+$ a maximum is apparent at n=13, and main
slope changes occur at n=4, 6, and 9.
Those features correlate with the observed abundance maxima.
The most prominent observed maximum \cite{Bar81,Cam81,Bar82,Ens83,Pfl86,Twu90}
is n=13.
The magic numbers at
n=4, 6 and 9, and specifically the enhanced abundances of
(NaI)$_6$Na$^+$ and (NaI)$_9$Na$^+$ clusters,~\cite{Twu90} are less
pronounced.
The only discrepancy between experiment and theoretical predictions occurs
for (NaI)$_9$Na$^+$. However,
the slope changes in Fig. 3 are so weak that the average binding energies
are not the best indicators of the enhanced stability of some
magic clusters.

The quantity $E_{bind}$ measures the cluster stability with respect to the
infinitely separated ions. The experiments indicate, however,
that the abundance mass spectrum should be probably best explained in terms of
the stability against evaporation of an alkali halide molecule.
\cite{Ens83,Twu90,Phi91}
The energy required to remove
a molecule AI from an (AI)$_n$A$^+$ cluster ion (A = Na, Cs)
is given by
\begin{eqnarray}
\label{disoci:energy}
E_{evaporation}
&=&
             E_{clus}[(AI)_{n-1}A^+] + E(AI)\\
\nonumber
&&
    - E_{clus}[(AI)_nA^+]
\end{eqnarray}
The evaporation energies are plotted in Fig. 4.
A sharp increase in the evaporation energy between n=14 and n=13, between
n=10 and n=9, between n=7 and n=6, and finally between n=5 and n=4 is
evident for the (CsI)$_n$Cs$^+$ clusters. This means that evaporative
cooling will result in enrichment of clusters with n=4, 6, 9 and 13 in the
beam. The results are similar for (NaI)$_n$Na$^+$, predicting enrichment of
clusters with n=4, 6 and 13, but a discrepancy with respect to experiment is
again obtained since enrichment is predicted for n=10, instead of n=9. In an
attempt to resolve this discrepancy we have also plotted in Fig. 4 a
``vertical'' evaporation energy. This is defined by removing from the
parent cluster (size n) the least-bound molecule (this can be identified in
the PI model, since the total binding energy of the cluster can be separated
into a sum of ion contributions; see Refs. \onlinecite{Lua90a,Ayuel,Agu97a,Agu97c}
for details), and relaxing the resulting structure (size n-1) to its nearest
local minimum. This is in many cases not the ground state of the (n-1)-cluster
and the difference between the ``adiabatic'' and ``vertical'' evaporation
energies in Fig. 4 accounts for this fact. In spite of this difference the
use of vertical evaporation energies leads to the same predictions for
(CsI)$_n$Cs$^+$ as before, but changes the predictions for (NaI)$_n$Na$^+$
to improve agreement with experiment for n=9, which is a maximum in the curve
of the vertical evaporation energies. The interpretation is that, although
the adiabatic evaporation of a molecule from (NaI)$_{10}$Na$^+$ costs more
energy than adiabatic evaporation from (NaI)$_9$Na$^+$, there are in both
cases isomeric forms of the (n-1)-clusters with i) a structure similar to
that of the n-cluster and ii) large energy barriers between those isomeric
forms and the ground state structure of the (n-1)-cluster, such that the
vertical evaporation from (NaI)$_9$Na$^+$ is larger. In summary, our
calculations suggest the possible relevance of isomers of the (n-1)-cluster
with a structure similar to that of the GS of the n-cluster to explain the
details of the mass spectra when evaporative cooling is involved. This point
deserves further investigation.
The main magic numbers n=4 and n=13 are a consequence of the
enhanced stability of very symmetrical rocksalt structures, in two and three
dimensions, respectively. On the other hand, n=6 and n=9
are ``fine structure'' peaks of the spectra and the explanation in terms of
structural features is less evident.
These occur for (CsI)$_n$Cs$^+$ because structures are formed that
optimize the value of the Madelung energy more efficiently than for neighboring
cluster sizes. 
(CsI)$_9$Cs$^+$ has some highly coordinated ions: one cation with coordination
6 and three anions with coordination 5. (CsI)$_6$Cs$^+$ also contains one
cation with coordination 6. At the same time the lowest coordination found in
these two clusters is 3. In contrast, some of the neighbor clusters, like
(CsI)$_5$Cs$^+$ and (CsI)$_{10}$Cs$^+$ contain some ions with coordination 2.
Those highly coordinated structures are less competitive for (NaI)$_6$Na$^+$
and (NaI)$_9$Na$^+$.
Figure 3(a) also shows, for comparison, the binding energies per ion of
neutral stoichiometric (NaI)$_n$ clusters. \cite{Agu97a} The local maxima
occur for n=6, 9 and 12, and have been associated to the formation of
compact structures with large atomic coordination compared to clusters with
n+1 and n-1 molecules.

\subsection{Ionization Energies and Structure}

In previous studies of alkali-halide clusters,\cite{Ayuel,Agu97a} we
have analyzed the variation of the ionization potential (IP) with the
cluster size. The vertical IP was calculated in the Koopmans'
approximation as the binding energy of the lowest bound electron in
the cluster, which is of course
located on a specific anion.
Here we investigate the relation between the geometrical structure and the
spectrum of electronic states for different isomers of the same cluster. This
could provide a way to distinguish isomers, already explored for other types
of clusters.~\cite{Bin95}
In the PI model the electrons near the HOMO level are localized on distinct
anions, and the different eigenvalues arise from the different atomic
environment around nonequivalent anions.
Thus, the set of one-electron energy eigenvalues characteristic 
of each isomer can be 
considered as a fingerprint of its structural shape.\cite{Mar83} As an example,
we present in Fig. 5 the orbital energy spectrum corresponding
to the two isomers of (NaI)$_{13}$Na$^+$ given in Fig. 1. It is apparent
that the two isomers have quite 
different spectra.
These could be measured by photoelectron spectroscopy,
and in principle it could be possible to determine the structure
of the different isomers present in a mass-selected beam by comparing the
experimental spectra with theoretical results. 
In our example, only two peaks are expected in the lowest energy part of
the spectrum for the case of the GS isomer,
because the symmetry of this structure induces high degeneracies.
One of the peaks corresponds to removing one electron from any of the twelve
surface anions (labelled A in the figure)
and is identified with the vertical IP. The other corresponds 
to the removal of one electron from 
the central anion (labelled B) with coordination six. The second isomer
has a larger number of inequivalent anions and the spectrum is broader.
Besides, the
ionization energy is 1.3 eV lower than for the GS
isomer. Similar ``fingerprints'' distinguish different isomers for other
cluster sizes.

\section{Summary}
\label{sec:summary}

The determination of the structures of alkali halide cluster isomers is a
challenging subject for present-day experimental techniques.\cite{Mai97,Hud97}
Theoretical calculations can throw light on these problems.
In this paper, we present {\em ab initio} calculations of the
structures and stabilities of (NaI)$_n$Na$^+$ and (CsI)$_n$Cs$^+$ 
cluster ions with (n$\le$14).
Starting from several local minima found with phenomenological
pair potential models, we have used the Perturbed Ion Model
(with correlation included) in order to determine
the ground state and some low-lying isomers.
Our results indicate an early formation of rocksalt fragments.
The rocksalt features appear at values of n lower than in the
corresponding neutral species.\cite{Agu97a,Agu97c}
Arguments related to the compactness of some clusters appear to
be able to explain the main features obtained in the mobility measurements
of Kreisle and coworkers.\cite{Mai97}
The mass spectra obtained by several experimental techniques
\cite{Bar81,Cam81,Bar82,Ens83,Pfl86,Twu90} show evidence
of enhanced population for cluster sizes n= 4, 6, 9, and 13. Our calculations
confirm the enhanced
stability of these clusters, namely, those
clusters are very stable against evaporation of a single molecule.
We have
investigated the possibility to determine isomeric structures by comparing
experimental photoelectron spectra with those obtained
theoretically. As an example we have shown how the spectra of
orbital energy eigenvalues of
two (NaI)$_{13}$Na$^+$ isomers depend on the structure. 
With all these results in mind, we are
confident that the identification of the isomer structures of
clusters of ionic materials can be feasible in a near future 
if experimental and
theoretical efforts work together.

$\;$

$\;$

$\;$

$\;$

{\bf ACKNOWLEDGMENTS}: Work supported by DGES (PB95-0720-C02-01) and Junta de
Castilla y Le\'on (VA63/96). A. Aguado is supported by a predoctoral fellowship
from Junta de Castilla y Le\'on. A. Ayuela acknowledges a Marie Curie
Fellowship supported by the EU TMR program.

\newpage
 
{\bf Captions of figures}

$\;$

{\bf Figure 1}. Lowest-energy structure and low-lying isomers of (NaI)$_n$Na$^+$.
The energy
difference (in eV) with respect to the most stable structure is given below
the corresponding isomers.

$\;$

{\bf Figure 2}. Lowest-energy structure and low-lying isomers of (CsI)$_n$Cs$^+$.
The energy
difference (in eV) with respect to the most stable structure is given below
the corresponding isomers.

$\;$

{\bf Figure 3}. Binding energy per ion as a function of the cluster size
of (NaI)$_n$Na$^+$ (circles) and (NaI)$_n$ (squares) in panel (a) and
of (CsI)$_n$Cs$^+$ in (b).

$\;$

{\bf Figure 4}.
Adiabatic (circles) and vertical (squares) energies required to 
evaporate a neutral molecule from
(NaI)$_n$Na$^+$ (a) and (CsI)$_n$Cs$^+$ (b) clusters as a function of n.

$\;$

{\bf Figure 5}.
Spectrum of orbital energy eigenvalues for the two isomers of
(NaI)$_{13}$Na$^+$ studied. 
Inequivalent anions are labelled with different letters. The vertical scale
gives the number of equivalent anions of each type.

\newpage

%\begin{figure}
%\vspace{-10mm}
%\psfig{figure=figg1.ps,width=85mm,height=115mm,rheight=75mm}
%%\caption{Screened pseudopotentials within BP GGA and LDA for germanium
%%and copper.  On the scale of these plots the GGA and LDA pseudopotentials
%%lie one on top of each other.  Shown are Troullier-Martins pseudopotentials
%%with cutoff radii $r_{s,p}=1.9$~bohr and $r_{d}=2.3$~bohr (Ge),
%%and $r_{s,d}= 2.0$~bohr and $r_{p}=2.3$~bohr (Cu).
%%}
%\label{fig:naina:isomers}
%\end{figure}

\end{document}